\documentclass[twocolumn,preprintnumbers,amsmath,amssymb,amsfonts]{revtex4-1}
\usepackage{epsfig}
\usepackage{graphicx}
\usepackage{subfigure}

\begin{document}
\preprint{UTTG-14-10}
\preprint{TCC-033-10}

\title{Dark Radiation Emerging After Big Bang Nucleosynthesis?}

\author{Willy Fischler}
\email{fischler@physics.utexas.edu}
\author{Joel Meyers}
\email{joelmeyers@mail.utexas.edu}
\affiliation{Theory Group, Department of Physics, University of Texas, Austin, TX 78712}
\affiliation{Texas Cosmology Center, University of Texas, Austin, TX 78712}

\date{\today}

\begin{abstract}
We show how recent data from observations of the cosmic microwave background may suggest the presence of additional radiation density which appeared after big bang nucleosynthesis.  We propose a general scheme by which this radiation could be produced from the decay of non-relativistic matter, we place constraints on the properties of such matter, and we give specific examples of scenarios in which this general scheme may be realized.
\end{abstract}

\maketitle

\section{Introduction}
The era that follows big bang nucleosynthesis and precedes matter domination is usually assumed to be dominated by radiation \cite{Weinberg:2008zzc}. The relativistic degrees of freedom during that era are also posited, by continuity, to be the same as those that prevailed during big bang nucleosynthesis.  However recent measurements by WMAP \cite{Komatsu:2010fb} and ACT \cite{Dunkley:2010ge} show a central value for the number of relativistic degrees of freedom during this era to be higher than the one inferred from big bang nucleosynthesis. With current data, the difference is not yet statistically significant. In this paper we will take as a working assumption that the central value reflects an accurate measurement of the relativistic energy density during that epoch. 

There are stringent bounds from big bang nucleosynthesis, particle physics, and the cosmic microwave background on the number and type of light degrees of freedom present in the early universe.  These bounds constrain the possible degrees of freedom that might be responsible for the new data. Attempts have been made to accommodate the new data by introducing relativistic sterile neutrinos \cite{Hamann:2010bk}, or by allowing for neutrino asymmetry \cite{Krauss:2010xg,Mangano:2010ei}.  We will develop in this paper an alternative explanation by postulating the existence of a degree of freedom that behaves as non-relativistic matter that is neutral under the visible gauge group $ SU(3)_C \times SU(2) \times U(1)$. This degree of freedom can be a non-relativistic particle or possibly a coherent oscillating (pseudo)scalar field, it contributes a fraction of the  energy density during big bang nucleosynthesis, and decays into dark radiation after light nuclei have been formed.  The resulting radiation adds a relativistic contribution to the energy density during the era of interest. 

If future data indeed produces a statistically significant difference between the effective number of neutrinos during big bang nucleosynthesis and the number obtained from cosmic microwave background observations, then this will pose a significant challenge to the sterile neutrino and neutrino asymmetry proposals in their present form.

The paper is organized as follows: Section \ref{a} will describe the relevant observational data which constrain the radiation content of the universe.  In Section \ref{b}, the focus is on the implications for the cosmic microwave background from the presence of an invisible non-relativistic particle decaying after big bang nucleosynthesis into dark radiation. We will discuss various regions of parameters, constrain the properties of the decay products, and comment about possible shortcomings of the sterile neutrino or neutrino asymmetry proposal.  In Section \ref{c}, we present two specific examples where this general proposal can be realized. We close with conclusions in Section \ref{d}.

\section{Observational Constraints}\label{a}

It has long been known that primordial light element abundances can be used to constrain the energy density of relativistic degrees of freedom during the period of big bang nucleosynthesis \cite{Fields:2006ga}.  More recently, observations of temperature fluctuations in the cosmic microwave background have begun to provide an independent constraint on the radiation energy density at a much later period in the history of the universe.  Specifically, for a fixed matter density, increasing the radiation density will result in a later time for matter-radiation equality, and this has the effect enhancing the early-time integrated Sachs-Wolfe effect.  Additionally, relativistic particles which are not coupled to the photon-baryon fluid are capable of free-streaming out of gravitational potential wells faster than the sound speed which provides a phase shift and damping of the acoustic peaks \cite{Bashinsky:2003tk}.

The bounds on radiation energy density are usually given in terms of a parameter $N_{eff}$ describing the effective number of neutrino species defined by
\begin{equation}\label{Neff}
	\rho_R = \left[1+N_{eff}\left(\frac{7}{8}\right)\left(\frac{4}{11}\right)^{4/3}\right]\rho_{\gamma}\, .
\end{equation}
The number of light neutrino species is constrained by measurements of the decay width of the $Z$ boson \cite{Nakamura:2010zzi} to be $2.984\pm0.008$.  The standard model prediction with three flavors of neutrinos and allowing for residual heating of the neutrino fluid due to electron-positron annihilation and other subtle effects \cite{Mangano:2005cc} is $N_{eff}=3.046$.  Observations of a value for $N_{eff}$ which is different from this prediction imply a departure from the standard scenario of a nearly thermal distribution of three flavors of standard model neutrinos.

Observations of primordial ${}^4$He abundance provide the best constraint on $N_{eff}$ during big bang nucleosynthesis.  These observations are consistent with the standard model predictions giving a constraint (at 68\% CL) \cite{Simha:2008zj}
\begin{equation}\label{BBN}
	N_{eff}^{BBN}=2.4\pm0.4\, .
\end{equation}
Some authors have recently reported a higher value for the primordial helium abundance \cite{Izotov:2010ca,Aver:2010wq}. The authors of \cite{Izotov:2010ca} argue that this may imply a larger value for $N_{eff}^{BBN}$, while the authors of \cite{Aver:2010wq} argue that the uncertainties are too large to suggest any tension with the standard big bang nucleosynthesis scenario.  This is an important issue which warrants further research, however, for the remainder of this paper, we will be using the constraint quoted above which seems to be consistent with earlier studies  \footnote{These observational contraints may be affected if one allows for non-standard neutrino physics.  See \cite{Dolgov:2002wy} for a review of some of these effects.}.

The current constraints from observations of the cosmic microwave background are significantly weaker than those from big bang nucleosynthesis.  The WMAP satellite \cite{Komatsu:2010fb} derives its constraints on $N_{eff}$ primarily from the first and third acoustic peaks, while the ground-based Atacama Cosmology Telescope \cite{Dunkley:2010ge} uses observations of the third through the seventh peaks.  These complementary measurements which span a broad range of scales give some confidence that the results of the measurements are not due to a feature which affects only a portion of angular power spectrum.  The current constraints are (at 68\% CL)
\begin{align}
	\mathrm{(WMAP7+BAO+{\it H_0})} \quad & N_{eff}^{CMB}=4.34_{-0.88}^{+0.86}\, ,\\
	\mathrm{(ACT+BAO+{\it H_0})} \qquad & N_{eff}^{CMB}=4.56 \pm 0.75\, .
\end{align}
The constraints listed here assumed $\mathrm{\Lambda CDM}$ cosmology with massless neutrinos.  Allowing for non-zero neutrino masses or varying dark energy equation of state affects the allowed parameter range \cite{Hamann:2010bk,Hamann:2010pw}.

These limits are consistent with $N_{eff}^{BBN}$ at the 2-$\sigma$ level, but there is a hint of tension between the central values of these measurements.  The Planck satellite will significantly increase the precision of cosmic microwave background observations \cite{Perotto:2006rj} giving $\delta N_{eff}^{CMB}\simeq0.26$ or better \cite{Hamann:2007sb}, and it will also help to break degeneracies with non-zero neutrino masses and varying equation of state for dark energy \cite{Hamann:2010pw}.  If the central values from the current measurements reflect an accurate measurement of $N_{eff}$, Planck could reveal a disagreement between $N_{eff}^{BBN}$ and $N_{eff}^{CMB}$ at the level of 4- to 5-$\sigma$.

\section{Decaying Matter}\label{b}

If Planck reveals a significant discrepancy between $N_{eff}^{BBN}$ and $N_{eff}^{CMB}$, it is important to ask how these results may be reconciled.  We suggest that the decay of non-relativistic matter into radiation would provide the necessary increase in $N_{eff}$ to resolve the conflict.  The energy density in the relativistic decay products, and hence the contribution to $N_{eff}^{CMB}$, will be determined by the energy density in the non-relativistic matter at some earlier time along with the lifetime of the matter.  For matter with an energy density $\rho_X$ at $t=10^{-4}$ s, and a lifetime $\tau$, we can calculate the increase in $N_{eff}$ after the decay to be
\begin{equation}\label{matter}
\Delta N_{eff}=\left(\frac{8}{7}\right)\left(\frac{11}{4}\right)^{4/3}\left(\frac{\tau}{10^{-4}\rm{s}}\right)^{1/2}\frac{\rho_X[t=10^{-4}\rm{s}]}{\rho_0}
\end{equation}
where $\rho_0$ is the photon energy density at $t=10^{-4}$ s adjusted to include the effect of heating due to electron positron annihilation, and is given by $\rho_0=a_{\mathcal{B}}(1.334\times10^{12}\rm{K})^4=1.149\times10^{-4} \mathrm{GeV}^4$ \cite{Weinberg:2008zzc}.  In deriving (\ref{matter}), we have used the fact that the energy density of non-relativistic matter decreases as $a^{-3}$ and that of radiation as $a^{-4}$, where $a$ is the Robertson-Walker scale factor which scales as $t^{1/2}$ during the radiation dominated era.  We have also assumed a very rapid decay into radiation at the time $t=\tau$.

\begin{figure}[!htp]
  \centering
   \subfigure[{\small WMAP 7-Year Constraints}]{\includegraphics[width=\columnwidth]{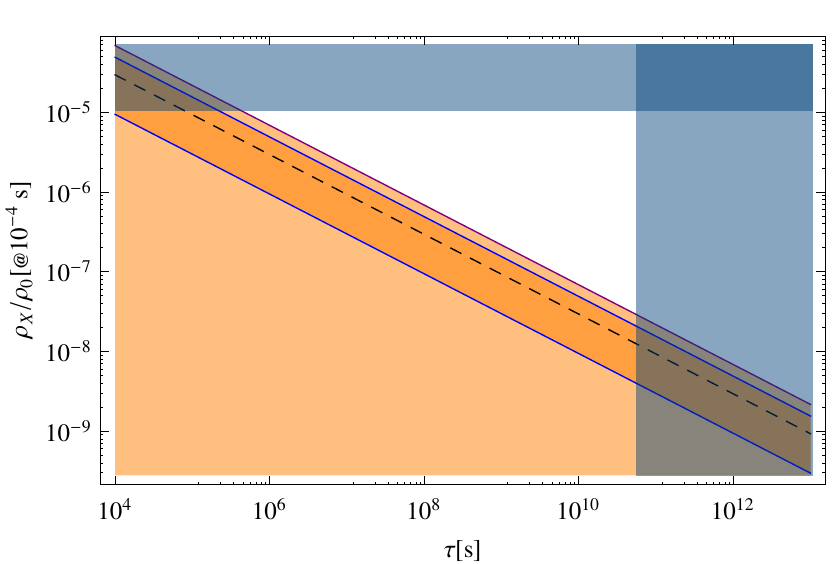}}                
   \subfigure[{\small Planck Projected Constraints}]{\includegraphics[width=\columnwidth]{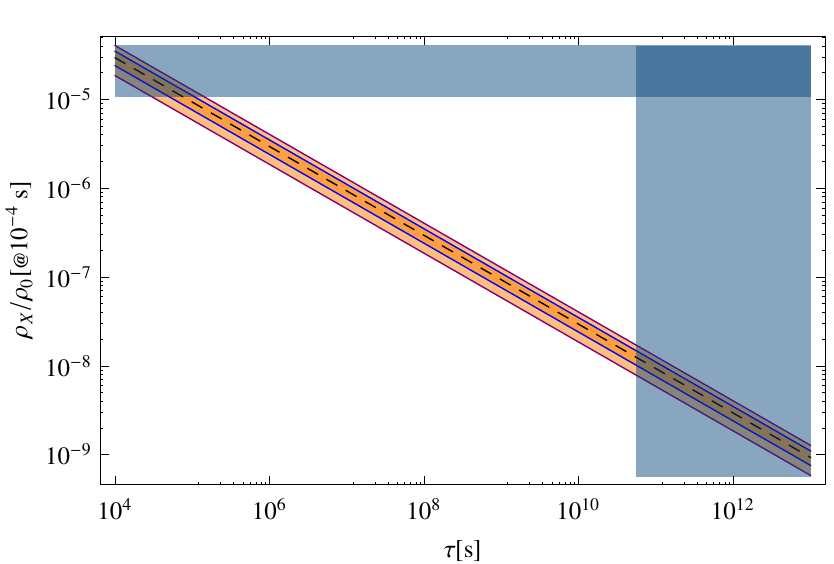}}
   \caption{\footnotesize These figures illustrate the constraints on energy density and lifetime of non-relativistic matter decaying into radiation.  The vertical axis gives the ratio of the energy density with $\rho_0 \equiv a_{\mathcal{B}}(1.334\times10^{12}\rm{K})^4=1.149\times10^{-4} \mathrm{GeV}^4$ evaluated at $10^{-4}$ seconds, and the horizontal axis shows the lifetime reported in seconds.  The blue shaded region near the top of the figure is excluded due to the constraint that $N_{eff}<3.2$ during big bang nucleosynthesis while the shaded region near the right is excluded by requiring that the decay into radiation occurs before the highest $l$ modes observable in the cosmic microwave background reenter the horizon.  The dark and light orange shaded regions give the 1- and 2-$\sigma$ constraints, and the black dashed line corresponds to the current observed central value of $N_{eff}^{CMB}=4.34$ from WMAP.  The constraints from WMAP 7-year data are consistent with no extra energy density, while the projected constraints from Planck require a nonzero energy density and fix the lifetime of the decaying matter.}
  \label{fig:decayplot}
\end{figure}

It is important that this additional matter give at most a subdominant contribution to the total energy density during the period of big bang nucleosynthesis so as not to alter the predictions of primordial light element abundances.  We thus require that $N_{eff}<3.2$ for $t<20$ minutes.  For a more detailed study of the constraints on decaying particles during big bang nucleosynthesis see \cite{Scherrer:1987rr,Scherrer:1987rs}.  It is also necessary that the decay into radiation takes place before the observable modes of the cosmic microwave background have reentered the horizon.  This requirement constrains $\tau<1650$ years \footnote{The maximum lifetime was determined by using CAMB \cite{Lewis:1999bs} which we modified to include the effects of non-relativistic matter decaying into radiation.}.  Putting this together, we can find the allowed range of parameters which satisfies the constraints from both $N_{eff}^{BBN}$ and $N_{eff}^{CMB}$.  We summarize our results in Figure \ref{fig:decayplot}.

It seems necessary that the relativistic decay products be neutral under the standard model gauge group.  Light particles carrying strong or weak charge are very well constrained by collider experiments \cite{Nakamura:2010zzi}.  A decay into photons would run the risk of destroying the nuclei produced during big bang nucleosynthesis, and thus altering the primordial light element abundances.  Furthermore, these photons would equilibrate in the photon-baryon plasma, increasing the photon temperature relative to the neutrino temperature, and therefore decreasing rather than increasing $N_{eff}$ as measured in the cosmic microwave background. 
This means that if a scenario such as the one we have described is realized in nature, it is unlikely that we would ever be able to observe the effects of the new contribution to the energy density, except through its gravitational signature.

A suggestion similar to the one made here appeared previously in the literature which was motivated by data from large scale structure and Lyman-$\alpha$ forest measurements \cite{Ichikawa:2007jv}.  The cosmological implications of that model were discussed in detail in \cite{Kawasaki:2007mk}.

We have focused on the case where the energy density before the decay is in the form of non-relativistic matter, however, this is not the only possibility.  We require only that there is some component of the energy density which redshifts more slowly than radiation so that the ratio $\rho_X/\rho_{\gamma}$ increases with time.  This allows $\rho_X$ to be subdominant during big bang nucleosynthesis while it may become important before the modes relevant for the cosmic microwave background have reentered the horizon.  Note that scenarios such as the addition of sterile neutrinos \cite{Hamann:2010bk}, other decoupled radiation \cite{Nakayama:2010vs}, or neutrino asymmetry \cite{Krauss:2010xg,Mangano:2010ei} tend to give $N_{eff}^{BBN}=N_{eff}^{CMB}$, and thus if future data reveals a statistically significant difference between $N_{eff}^{BBN}$ and $N_{eff}^{CMB}$, these models would face a significant challenge.  The requirements on the decay products are more restrictive because they must act as radiation in order to have the same effect as an increase in $N_{eff}$.  These particles need not be strictly massless, but it is important that they are relativistic until well after matter-radiation equality so as not to significantly alter late time cosmology \footnote{For a discussion of constraints on cosmological fluids decaying into radiation at late times, see \cite{Dutta:2009ix,Dutta:2010kb,Doroshkevich:1989bf}}.  For a fluid component with a pressure given by $p_X=w\rho_X$ with $w<1/3$, equation (\ref{matter}) becomes
\begin{equation}\label{generalw}
\Delta N_{eff}=\left(\frac{8}{7}\right)\left(\frac{11}{4}\right)^{4/3}\left(\frac{\tau}{10^{-4}\rm{s}}\right)^{(1-3w)/2}\frac{\rho_X[t=10^{-4}\rm{s}]}{\rho_0}\, .
\end{equation}

\section{Examples}\label{c}
\subsection{Non-Relativistic Invisible Particle}\label{nrip}

A context in which such non-relativistic particles occur naturally are supersymmetric models which exhibit a hierarchy of splittings in super-multiplets \cite{Fischler:2010nk}. This hierarchy was exemplified in \cite{Fischler:2010nk} in the context of gauge mediation \cite{Dine:1981za,Dimopoulos:1981au,Nappi:1982hm,AlvarezGaume:1981wy,Dine:1981gu,Dine:1982zb,Dine:1993yw,Dine:1994vc,Dine:1995ag,Meade:2008wd} with mediators being charged under the standard gauge group as well as an invisible gauge group. The hierarchy in masses of the mediator fields reflects itself in the hierarchy noted in splittings. Details of the construction can be found in the paper by Fischler and Tangarife-Garcia \cite{Fischler:2010nk}.  For other models with decaying particles see for example \cite{Dolgov:2002wy,Bowen:2001in,Cuoco:2005qr,Ichikawa:2007jv,Kawasaki:2007mk}.

 A prime candidate as shown in the aforementioned paper is an invisible photino. This particle is neutral under the visible gauge group,
 $ SU(3)_C \times SU(2) \times U(1)$ and is the super-partner of an invisible $U(1)$ gauge field. The low mass of this particle is a consequence of the small splitting inherited by the super-multiplets in the invisible sector. The invisible photino decays into an invisible photon and a relativistic gravitino. It's lifetime, cross-sections, and mass are related to various parameters in the class of models mentioned above. These parameters include in particular, the supersymmetry breaking scale, the mass of the relevant mediators, and gauge coupling constants. The abundance of the invisible photino is fixed by the property  that it is in equilibrium after reheating with other invisible degrees of freedom including invisible photons. As noted in \cite{Fischler:2010nk} the temperature of the invisible sector is lower than the visible sector.  Below, we exhibit the values for quantities related to parameters occurring in a specific model belonging to this class and adjusted so that they are in agreement with the values obtained from Figure \ref{fig:decayplot}. An additional requirement is that the gravitino mass not affect late time cosmology.
 
 In particular, we consider three parameters: the SUSY breaking scale $\mu$, the photino mass $m_{\tilde{\gamma}}$, and the annihilation cross section $\langle \sigma v \rangle$. These are given as functions of $\tau$ by the following expressions.
\begin{align}
m_{\tilde{\gamma}}(\tau)\,&=\,(5.049\times10^{-1}\,\mathrm{GeV})\beta^{5/16}\left(\frac{1{\rm s}}{\tau}\right)^{1/4}\Delta N_{eff}^{-1/4}\, , \\
\mu (\tau) \,&=\,(4.573\times10^6\,\mathrm{GeV})\beta^{5/6}\left(\frac{1{\rm s}}{\tau}\right)^{1/16}\Delta N_{eff}^{-5/16}\, , \\
\langle \sigma v \rangle (\tau)\,&=\,\frac{x_f}{2.098\times10^{15}\,\mathrm{GeV}^2}\left(\frac{1{\rm s}}{\tau}\right)^{-1/2}\Delta N_{eff}^{-1}\, ,
\end{align}
where $\beta$ is the ratio between the current radiation energy density and the current gravitino energy density \[ \beta = \frac{\rho_{\tilde{G},0}}{\rho_{\gamma,0}}.\]  and $x_f = m_{\tilde{\gamma}}/T_f$ is the ``freeze-out" parameter which is commonly taken to be $x_f\sim 20$. For the specific model presented in the Fischler-Tangarife paper, these expressions can be rewritten in terms of a new set of parameters $m_2$, $\mu$ and $v_h$ by using the relations:
\begin{align}
 m_{\tilde{\gamma} }\,&\approx\,\frac{\alpha_h }{4\pi}\frac{\mu^2}{m_2}\, , \\
\langle\sigma v \rangle &\sim \frac{\alpha_h m^4_{\tilde{\gamma}}}{4\pi^3v_h^6}\, .
\end{align}

\subsection{\label{osc} \bf Oscillating (Pseudo)Scalar}
 
 Another possible realization of our generic proposal is to consider a (pseudo)scalar field whose homogeneous coherent oscillations mimic non-relativistic matter and contribute a sub-dominant component of the energy density. This (pseudo)scalar field decays into dark photons via the following dimension 5 operator: 
\begin{equation}\label{phif2}
  \frac{\phi (F_{\mu \nu} )^2}{M}\, .
\end{equation}
The potential for this (pseudo)scalar field is assumed to be harmonic in the region of field space where the coherent (pseudo)scalar field is located at the onset of its oscillations.
\begin{equation}\label{V}
V= m^2\phi^2
\end{equation}

In a nutshell, the (pseudo)scalar field starts oscillating when the Hubble constant is of order of the mass of the (pseudo)scalar field. The lifetime can be obtained from the dimension 5 operator mentioned above. It is then straightforward to find how the energy density of the background at the time of decay, $\rho{_\tau}$ depends on the energy density of the background at the onset of oscillations $\rho{_0}$.
\begin{equation}
 \rho{_\tau} \sim \rho{_0} \left(\frac{m}{M}\right)^3 =  {m^2 {\phi_0}^2} \left(\frac{m}{M}\right)^3 
\end{equation}

The above generic constraints then translate into restrictions on the value of $M, m, \phi_0$, where $M$ is the denominator mass of the dimension 5 operator, $m$ is the mass of the (pseudo)scalar field, and $\phi_0$ is the value of $\phi$ at the onset of oscillations.

The parameters are given by the following expressions
\begin{align}
m_\phi(\tau)\,&=\, (2.417\times10^{-27}\,\mathrm{GeV})\xi^{-4}\left(\frac{1{\rm s}}{\tau}\right) \Delta N_{eff}^2 \\
%
%
%
M(\tau)\,&=\,(1.188\times10^{-28}\,\mathrm{GeV})\xi^{-6}\left( \frac{1{\rm s}}{\tau} \right)\Delta N_{eff}^3\, ,
\end{align}
where $\xi$ defined by
\begin{equation}
	\phi_0\,=\,\xi M_P \, , \nonumber
\end{equation}
can take values in the range 
\begin{equation}
	3\times 10^{-9} < \xi < 10^{-7} \, . \nonumber
\end{equation}  

For the upper bound of $\xi$ ($\xi\sim 10^{-7}$ and using $N_{eff}\sim 4.6$), the minimum values  that can be obtained for $m_\phi$ and $M$ are: 
\begin{equation}
 m_\phi\sim 10^{-1} \,{\rm eV}, \,\,\,\,M\sim 10^3 \,{\rm GeV} \, . \nonumber
\end{equation}

For the lower bound, $\xi\sim 3\times 10^{-9}$, we get that the maximum values for these masses are:
\begin{equation}
 m_\phi\sim 10^{2} \,{\rm GeV}, \,\,\,\,M\sim 10^{18} \,{\rm GeV} \, . \nonumber
\end{equation}

\section{Conclusions}\label{d}

If one assumes the effective number of neutrino species before and after big bang nucleosynthesis is indeed accurately given by the central values of the experimental data, a conundrum ensues requiring the existence of a new kind of dark radiation which appears after the formation of light nuclei. With this assumption current realizations of sterile neutrinos and neutrino asymmetries cannot account for the difference between $N_{eff}^{BBN}$ and $N_{eff}^{CMB}$, however with the present experimental precision these scenarios are consistent within the experimental errors.  We have shown how the decay of non-relativistic matter into dark radiation after big bang nucleosynthesis may reconcile a difference in $N_{eff}^{BBN}$ and $N_{eff}^{CMB}$ if forthcoming data indeed reveals such a difference.  Our results for the constraints on the energy density and lifetime of such matter are summarized in Figure \ref{fig:decayplot}.  We described two specific examples in which this general scenario could be realized: a supersymmetric model with a hierarchy of splittings in super-multiplets, and an oscillating (pseudo)scalar field.

\section{Acknowledgments}
The authors would like to thank Walter Tangarife-Garcia, Richard Leu, and Eiichiro Komatsu for very helpful discussions.
The research of W.F. and J. M. was supported in part by the National Science Foundation under Grant Numbers PHY-0969020 and PHY-0455649.

\bibliography{wj}

\begin{thebibliography}{10}%
\makeatletter
\providecommand \@ifxundefined [1]{%
 \ifx #1\undefined \expandafter \@firstoftwo
 \else \expandafter \@secondoftwo
\fi
}%
\providecommand \@ifnum [1]{%
 \ifnum #1\expandafter \@firstoftwo
 \else \expandafter \@secondoftwo
\fi
}%
\providecommand \enquote [1]{``#1''}%
\providecommand \bibnamefont  [1]{#1}%
\providecommand \bibfnamefont [1]{#1}%
\providecommand \citenamefont [1]{#1}%
\providecommand\href[0]{\@sanitize\@href}%
\providecommand\@href[1]{\endgroup\@@startlink{#1}\endgroup\@@href}%
\providecommand\@@href[1]{#1\@@endlink}%
\providecommand \@sanitize [0]{\begingroup\catcode`\&12\catcode`\#12\relax}%
\@ifxundefined \pdfoutput {\@firstoftwo}{%
 \@ifnum{\z@=\pdfoutput}{\@firstoftwo}{\@secondoftwo}%
}{%
 \providecommand\@@startlink[1]{\leavevmode\special{html:<a href="#1">}}%
 \providecommand\@@endlink[0]{\special{html:</a>}}%
}{%
 \providecommand\@@startlink[1]{%
  \leavevmode
  \pdfstartlink
   attr{/Border[0 0 1 ]/H/I/C[0 1 1]}%
   user{/Subtype/Link/A<</Type/Action/S/URI/URI(#1)>>}%
  \relax
 }%
 \providecommand\@@endlink[0]{\pdfendlink}%
}%
\providecommand \url  [0]{\begingroup\@sanitize \@url }%
\providecommand \@url [1]{\endgroup\@href {#1}{\urlprefix}}%
\providecommand \urlprefix [0]{URL }%
\providecommand \Eprint[0]{\href }%
\@ifxundefined \urlstyle {%
  \providecommand \doi [1]{doi:\discretionary{}{}{}#1}%
}{%
  \providecommand \doi [0]{doi:\discretionary{}{}{}\begingroup
  \urlstyle{rm}\Url }%
}%
\providecommand \doibase [0]{http://dx.doi.org/}%
\providecommand \Doi[1]{\href{\doibase#1}}%
\providecommand \bibAnnote [3]{%
  \BibitemShut{#1}%
  \begin{quotation}\noindent
    \textsc{Key:}\ #2\\\textsc{Annotation:}\ #3%
  \end{quotation}%
}%
\providecommand \bibAnnoteFile [2]{%
  \IfFileExists{#2}{\bibAnnote {#1} {#2} {\input{#2}}}{}%
}%
\providecommand \typeout [0]{\immediate \write \m@ne }%
\providecommand \selectlanguage [0]{\@gobble}%
\providecommand \bibinfo [0]{\@secondoftwo}%
\providecommand \bibfield [0]{\@secondoftwo}%
\providecommand \translation [1]{[#1]}%
\providecommand \BibitemOpen[0]{}%
\providecommand \bibitemStop [0]{}%
\providecommand \bibitemNoStop [0]{.\EOS\space}%
\providecommand \EOS [0]{\spacefactor3000\relax}%
\providecommand \BibitemShut [1]{\csname bibitem#1\endcsname}%
\bibitem{Weinberg:2008zzc}%
  \BibitemOpen
  \bibfield{author}{%
  \bibinfo {author} {\bibfnamefont{S.}~\bibnamefont{Weinberg}}\ }%
  \bibinfo {note} {oxford, UK: Oxford Univ. Pr. (2008) 593 p}%
  \bibAnnoteFile{NoStop}{Weinberg:2008zzc}%
\bibitem{Komatsu:2010fb}%
  \BibitemOpen
  \bibfield{author}{%
  \bibinfo {author} {\bibfnamefont{E.}~\bibnamefont{Komatsu}} \emph{et~al.}}%
   (\bibinfo {year} {2010}),\
  \Eprint{http://arxiv.org/abs/1001.4538}{arXiv:1001.4538 [astro-ph.CO]}%
  \bibAnnoteFile{NoStop}{Komatsu:2010fb}%
\bibitem{Dunkley:2010ge}%
  \BibitemOpen
  \bibfield{author}{%
  \bibinfo {author} {\bibfnamefont{J.}~\bibnamefont{Dunkley}} \emph{et~al.}}%
   (\bibinfo {year} {2010}),\
  \Eprint{http://arxiv.org/abs/1009.0866}{arXiv:1009.0866 [astro-ph.CO]}%
  \bibAnnoteFile{NoStop}{Dunkley:2010ge}%
\bibitem{Hamann:2010bk}%
  \BibitemOpen
  \bibfield{author}{%
  \bibinfo {author} {\bibfnamefont{J.}~\bibnamefont{Hamann}}, \bibinfo {author}
  {\bibfnamefont{S.}~\bibnamefont{Hannestad}}, \bibinfo {author}
  {\bibfnamefont{G.~G.}\ \bibnamefont{Raffelt}}, \bibinfo {author}
  {\bibfnamefont{I.}~\bibnamefont{Tamborra}},\ and\ \bibinfo {author}
  {\bibfnamefont{Y.~Y.~Y.}\ \bibnamefont{Wong}},\ }%
  \bibfield{journal}{%
  \Doi{10.1103/PhysRevLett.105.181301}{\bibinfo {journal} {Phys. Rev. Lett.}}\
  }%
  \textbf{\bibinfo {volume} {105}},\ \bibinfo {pages} {181301} (\bibinfo {year}
  {2010}),\ \Eprint{http://arxiv.org/abs/1006.5276}{arXiv:1006.5276 [hep-ph]}%
  \bibAnnoteFile{NoStop}{Hamann:2010bk}%
\bibitem{Krauss:2010xg}%
  \BibitemOpen
  \bibfield{author}{%
  \bibinfo {author} {\bibfnamefont{L.~M.}\ \bibnamefont{Krauss}}, \bibinfo
  {author} {\bibfnamefont{C.}~\bibnamefont{Lunardini}},\ and\ \bibinfo {author}
  {\bibfnamefont{C.}~\bibnamefont{Smith}}}%
   (\bibinfo {year} {2010}),\
  \Eprint{http://arxiv.org/abs/1009.4666}{arXiv:1009.4666 [hep-ph]}%
  \bibAnnoteFile{NoStop}{Krauss:2010xg}%
\bibitem{Mangano:2010ei}%
  \BibitemOpen
  \bibfield{author}{%
  \bibinfo {author} {\bibfnamefont{G.}~\bibnamefont{Mangano}}, \bibinfo
  {author} {\bibfnamefont{G.}~\bibnamefont{Miele}}, \bibinfo {author}
  {\bibfnamefont{S.}~\bibnamefont{Pastor}}, \bibinfo {author}
  {\bibfnamefont{O.}~\bibnamefont{Pisanti}},\ and\ \bibinfo {author}
  {\bibfnamefont{S.}~\bibnamefont{Sarikas}}}%
   (\bibinfo {year} {2010}),\
  \Eprint{http://arxiv.org/abs/1011.0916}{arXiv:1011.0916 [astro-ph.CO]}%
  \bibAnnoteFile{NoStop}{Mangano:2010ei}%
\bibitem{Fields:2006ga}%
  \BibitemOpen
  \bibfield{author}{%
  \bibinfo {author} {\bibfnamefont{B.}~\bibnamefont{Fields}}\ and\ \bibinfo
  {author} {\bibfnamefont{S.}~\bibnamefont{Sarkar}}}%
   (\bibinfo {year} {2006}),\
  \Eprint{http://arxiv.org/abs/astro-ph/0601514}{arXiv:astro-ph/0601514}%
  \bibAnnoteFile{NoStop}{Fields:2006ga}%
\bibitem{Bashinsky:2003tk}%
  \BibitemOpen
  \bibfield{author}{%
  \bibinfo {author} {\bibfnamefont{S.}~\bibnamefont{Bashinsky}}\ and\ \bibinfo
  {author} {\bibfnamefont{U.}~\bibnamefont{Seljak}},\ }%
  \bibfield{journal}{%
  \Doi{10.1103/PhysRevD.69.083002}{\bibinfo {journal} {Phys. Rev.}}\ }%
  \textbf{\bibinfo {volume} {D69}},\ \bibinfo {pages} {083002} (\bibinfo {year}
  {2004}),\
  \Eprint{http://arxiv.org/abs/astro-ph/0310198}{arXiv:astro-ph/0310198}%
  \bibAnnoteFile{NoStop}{Bashinsky:2003tk}%
\bibitem{Nakamura:2010zzi}%
  \BibitemOpen
  \bibfield{author}{%
  \bibinfo {author} {\bibfnamefont{K.}~\bibnamefont{Nakamura}} \emph{et~al.}
  (\bibinfo {collaboration} {Particle Data Group}),\ }%
  \bibfield{journal}{%
  \Doi{10.1088/0954-3899/37/7A/075021}{\bibinfo {journal} {J. Phys.}}\ }%
  \textbf{\bibinfo {volume} {G37}},\ \bibinfo {pages} {075021} (\bibinfo {year}
  {2010})%
  \bibAnnoteFile{NoStop}{Nakamura:2010zzi}%
\bibitem{Mangano:2005cc}%
  \BibitemOpen
  \bibfield{author}{%
  \bibinfo {author} {\bibfnamefont{G.}~\bibnamefont{Mangano}} \emph{et~al.},\
  }%
  \bibfield{journal}{%
  \Doi{10.1016/j.nuclphysb.2005.09.041}{\bibinfo {journal} {Nucl. Phys.}}\ }%
  \textbf{\bibinfo {volume} {B729}},\ \bibinfo {pages} {221} (\bibinfo {year}
  {2005}),\ \Eprint{http://arxiv.org/abs/hep-ph/0506164}{arXiv:hep-ph/0506164}%
  \bibAnnoteFile{NoStop}{Mangano:2005cc}%
\bibitem{Simha:2008zj}%
  \BibitemOpen
  \bibfield{author}{%
  \bibinfo {author} {\bibfnamefont{V.}~\bibnamefont{Simha}}\ and\ \bibinfo
  {author} {\bibfnamefont{G.}~\bibnamefont{Steigman}},\ }%
  \bibfield{journal}{%
  \Doi{10.1088/1475-7516/2008/06/016}{\bibinfo {journal} {JCAP}}\ }%
  \textbf{\bibinfo {volume} {0806}},\ \bibinfo {pages} {016} (\bibinfo {year}
  {2008}),\ \Eprint{http://arxiv.org/abs/0803.3465}{arXiv:0803.3465
  [astro-ph]}%
  \bibAnnoteFile{NoStop}{Simha:2008zj}%
\bibitem{Izotov:2010ca}%
  \BibitemOpen
  \bibfield{author}{%
  \bibinfo {author} {\bibfnamefont{Y.~I.}\ \bibnamefont{Izotov}}\ and\ \bibinfo
  {author} {\bibfnamefont{T.~X.}\ \bibnamefont{Thuan}},\ }%
  \bibfield{journal}{%
  \Doi{10.1088/2041-8205/710/1/L67}{\bibinfo {journal} {Astrophys. J.}}\ }%
  \textbf{\bibinfo {volume} {710}},\ \bibinfo {pages} {L67} (\bibinfo {year}
  {2010}),\ \Eprint{http://arxiv.org/abs/1001.4440}{arXiv:1001.4440
  [astro-ph.CO]}%
  \bibAnnoteFile{NoStop}{Izotov:2010ca}%
\bibitem{Aver:2010wq}%
  \BibitemOpen
  \bibfield{author}{%
  \bibinfo {author} {\bibfnamefont{E.}~\bibnamefont{Aver}}, \bibinfo {author}
  {\bibfnamefont{K.~A.}\ \bibnamefont{Olive}},\ and\ \bibinfo {author}
  {\bibfnamefont{E.~D.}\ \bibnamefont{Skillman}},\ }%
  \bibfield{journal}{%
  \Doi{10.1088/1475-7516/2010/05/003}{\bibinfo {journal} {JCAP}}\ }%
  \textbf{\bibinfo {volume} {1005}},\ \bibinfo {pages} {003} (\bibinfo {year}
  {2010}),\ \Eprint{http://arxiv.org/abs/1001.5218}{arXiv:1001.5218
  [astro-ph.CO]}%
  \bibAnnoteFile{NoStop}{Aver:2010wq}%
\bibitem{Note1}%
  \BibitemOpen
  \bibinfo {note} {These observational contraints may be affected if one allows
  for non-standard neutrino physics. See \cite {Dolgov:2002wy} for a review of
  some of these effects.}%
  \bibAnnoteFile{Stop}{Note1}%
\bibitem{Hamann:2010pw}%
  \BibitemOpen
  \bibfield{author}{%
  \bibinfo {author} {\bibfnamefont{J.}~\bibnamefont{Hamann}}, \bibinfo {author}
  {\bibfnamefont{S.}~\bibnamefont{Hannestad}}, \bibinfo {author}
  {\bibfnamefont{J.}~\bibnamefont{Lesgourgues}}, \bibinfo {author}
  {\bibfnamefont{C.}~\bibnamefont{Rampf}},\ and\ \bibinfo {author}
  {\bibfnamefont{Y.~Y.~Y.}\ \bibnamefont{Wong}},\ }%
  \bibfield{journal}{%
  \Doi{10.1088/1475-7516/2010/07/022}{\bibinfo {journal} {JCAP}}\ }%
  \textbf{\bibinfo {volume} {1007}},\ \bibinfo {pages} {022} (\bibinfo {year}
  {2010}),\ \Eprint{http://arxiv.org/abs/1003.3999}{arXiv:1003.3999
  [astro-ph.CO]}%
  \bibAnnoteFile{NoStop}{Hamann:2010pw}%
\bibitem{Perotto:2006rj}%
  \BibitemOpen
  \bibfield{author}{%
  \bibinfo {author} {\bibfnamefont{L.}~\bibnamefont{Perotto}}, \bibinfo
  {author} {\bibfnamefont{J.}~\bibnamefont{Lesgourgues}}, \bibinfo {author}
  {\bibfnamefont{S.}~\bibnamefont{Hannestad}}, \bibinfo {author}
  {\bibfnamefont{H.}~\bibnamefont{Tu}},\ and\ \bibinfo {author}
  {\bibfnamefont{Y.~Y.~Y.}\ \bibnamefont{Wong}},\ }%
  \bibfield{journal}{%
  \Doi{10.1088/1475-7516/2006/10/013}{\bibinfo {journal} {JCAP}}\ }%
  \textbf{\bibinfo {volume} {0610}},\ \bibinfo {pages} {013} (\bibinfo {year}
  {2006}),\
  \Eprint{http://arxiv.org/abs/astro-ph/0606227}{arXiv:astro-ph/0606227}%
  \bibAnnoteFile{NoStop}{Perotto:2006rj}%
\bibitem{Hamann:2007sb}%
  \BibitemOpen
  \bibfield{author}{%
  \bibinfo {author} {\bibfnamefont{J.}~\bibnamefont{Hamann}}, \bibinfo {author}
  {\bibfnamefont{J.}~\bibnamefont{Lesgourgues}},\ and\ \bibinfo {author}
  {\bibfnamefont{G.}~\bibnamefont{Mangano}},\ }%
  \bibfield{journal}{%
  \Doi{10.1088/1475-7516/2008/03/004}{\bibinfo {journal} {JCAP}}\ }%
  \textbf{\bibinfo {volume} {0803}},\ \bibinfo {pages} {004} (\bibinfo {year}
  {2008}),\ \Eprint{http://arxiv.org/abs/0712.2826}{arXiv:0712.2826
  [astro-ph]}%
  \bibAnnoteFile{NoStop}{Hamann:2007sb}%
\bibitem{Scherrer:1987rr}%
  \BibitemOpen
  \bibfield{author}{%
  \bibinfo {author} {\bibfnamefont{R.~J.}\ \bibnamefont{Scherrer}}\ and\
  \bibinfo {author} {\bibfnamefont{M.~S.}\ \bibnamefont{Turner}},\ }%
  \bibfield{journal}{%
  \Doi{10.1086/166534}{\bibinfo {journal} {Astrophys. J.}}\ }%
  \textbf{\bibinfo {volume} {331}},\ \bibinfo {pages} {19} (\bibinfo {year}
  {1988})%
  \bibAnnoteFile{NoStop}{Scherrer:1987rr}%
\bibitem{Scherrer:1987rs}%
  \BibitemOpen
  \bibfield{author}{%
  \bibinfo {author} {\bibfnamefont{R.~J.}\ \bibnamefont{{Scherrer}}}\ and\
  \bibinfo {author} {\bibfnamefont{M.~S.}\ \bibnamefont{{Turner}}},\ }%
  \bibfield{journal}{%
  \Doi{10.1086/166535}{\bibinfo {journal} {\apj}}\ }%
  \textbf{\bibinfo {volume} {331}},\ \bibinfo {pages} {33} (\bibinfo {month}
  {Aug.}\ \bibinfo {year} {1988})%
  \bibAnnoteFile{NoStop}{Scherrer:1987rs}%
\bibitem{Note2}%
  \BibitemOpen
  \bibinfo {note} {The maximum lifetime was determined by using CAMB \cite
  {Lewis:1999bs} which we modified to include the effects of non-relativistic
  matter decaying into radiation.}%
  \bibAnnoteFile{Stop}{Note2}%
\bibitem{Ichikawa:2007jv}%
  \BibitemOpen
  \bibfield{author}{%
  \bibinfo {author} {\bibfnamefont{K.}~\bibnamefont{Ichikawa}}, \bibinfo
  {author} {\bibfnamefont{M.}~\bibnamefont{Kawasaki}}, \bibinfo {author}
  {\bibfnamefont{K.}~\bibnamefont{Nakayama}}, \bibinfo {author}
  {\bibfnamefont{M.}~\bibnamefont{Senami}},\ and\ \bibinfo {author}
  {\bibfnamefont{F.}~\bibnamefont{Takahashi}},\ }%
  \bibfield{journal}{%
  \Doi{10.1088/1475-7516/2007/05/008}{\bibinfo {journal} {JCAP}}\ }%
  \textbf{\bibinfo {volume} {0705}},\ \bibinfo {pages} {008} (\bibinfo {year}
  {2007}),\ \Eprint{http://arxiv.org/abs/hep-ph/0703034}{arXiv:hep-ph/0703034
  [HEP-PH]}%
  \bibAnnoteFile{NoStop}{Ichikawa:2007jv}%
\bibitem{Kawasaki:2007mk}%
  \BibitemOpen
  \bibfield{author}{%
  \bibinfo {author} {\bibfnamefont{M.}~\bibnamefont{Kawasaki}}, \bibinfo
  {author} {\bibfnamefont{K.}~\bibnamefont{Nakayama}},\ and\ \bibinfo {author}
  {\bibfnamefont{M.}~\bibnamefont{Senami}},\ }%
  \bibfield{journal}{%
  \Doi{10.1088/1475-7516/2008/03/009}{\bibinfo {journal} {JCAP}}\ }%
  \textbf{\bibinfo {volume} {0803}},\ \bibinfo {pages} {009} (\bibinfo {year}
  {2008}),\ \Eprint{http://arxiv.org/abs/0711.3083}{arXiv:0711.3083 [hep-ph]}%
  \bibAnnoteFile{NoStop}{Kawasaki:2007mk}%
\bibitem{Nakayama:2010vs}%
  \BibitemOpen
  \bibfield{author}{%
  \bibinfo {author} {\bibfnamefont{K.}~\bibnamefont{Nakayama}}, \bibinfo
  {author} {\bibfnamefont{F.}~\bibnamefont{Takahashi}},\ and\ \bibinfo {author}
  {\bibfnamefont{T.~T.}\ \bibnamefont{Yanagida}}}%
   (\bibinfo {year} {2010}),\
  \Eprint{http://arxiv.org/abs/1010.5693}{arXiv:1010.5693 [hep-ph]}%
  \bibAnnoteFile{NoStop}{Nakayama:2010vs}%
\bibitem{Note3}%
  \BibitemOpen
  \bibinfo {note} {For a discussion of constraints on cosmological fluids
  decaying into radiation at late times, see \cite
  {Dutta:2009ix,Dutta:2010kb,Doroshkevich:1989bf}}%
  \bibAnnoteFile{NoStop}{Note3}%
\bibitem{Fischler:2010nk}%
  \BibitemOpen
  \bibfield{author}{%
  \bibinfo {author} {\bibfnamefont{W.}~\bibnamefont{Fischler}}\ and\ \bibinfo
  {author} {\bibfnamefont{W.~T.}\ \bibnamefont{Garcia}}}%
   (\bibinfo {year} {2010}),\
  \Eprint{http://arxiv.org/abs/1011.0099}{arXiv:1011.0099 [hep-ph]}%
  \bibAnnoteFile{NoStop}{Fischler:2010nk}%
\bibitem{Dine:1981za}%
  \BibitemOpen
  \bibfield{author}{%
  \bibinfo {author} {\bibfnamefont{M.}~\bibnamefont{Dine}}, \bibinfo {author}
  {\bibfnamefont{W.}~\bibnamefont{Fischler}},\ and\ \bibinfo {author}
  {\bibfnamefont{M.}~\bibnamefont{Srednicki}},\ }%
  \bibfield{journal}{%
  \Doi{10.1016/0550-3213(81)90582-4}{\bibinfo {journal} {Nucl. Phys.}}\ }%
  \textbf{\bibinfo {volume} {B189}},\ \bibinfo {pages} {575} (\bibinfo {year}
  {1981})%
  \bibAnnoteFile{NoStop}{Dine:1981za}%
\bibitem{Dimopoulos:1981au}%
  \BibitemOpen
  \bibfield{author}{%
  \bibinfo {author} {\bibfnamefont{S.}~\bibnamefont{Dimopoulos}}\ and\ \bibinfo
  {author} {\bibfnamefont{S.}~\bibnamefont{Raby}},\ }%
  \bibfield{journal}{%
  \Doi{10.1016/0550-3213(81)90430-2}{\bibinfo {journal} {Nucl. Phys.}}\ }%
  \textbf{\bibinfo {volume} {B192}},\ \bibinfo {pages} {353} (\bibinfo {year}
  {1981})%
  \bibAnnoteFile{NoStop}{Dimopoulos:1981au}%
\bibitem{Nappi:1982hm}%
  \BibitemOpen
  \bibfield{author}{%
  \bibinfo {author} {\bibfnamefont{C.~R.}\ \bibnamefont{Nappi}}\ and\ \bibinfo
  {author} {\bibfnamefont{B.~A.}\ \bibnamefont{Ovrut}},\ }%
  \bibfield{journal}{%
  \Doi{10.1016/0370-2693(82)90418-X}{\bibinfo {journal} {Phys. Lett.}}\ }%
  \textbf{\bibinfo {volume} {B113}},\ \bibinfo {pages} {175} (\bibinfo {year}
  {1982})%
  \bibAnnoteFile{NoStop}{Nappi:1982hm}%
\bibitem{AlvarezGaume:1981wy}%
  \BibitemOpen
  \bibfield{author}{%
  \bibinfo {author} {\bibfnamefont{L.}~\bibnamefont{Alvarez-Gaume}}, \bibinfo
  {author} {\bibfnamefont{M.}~\bibnamefont{Claudson}},\ and\ \bibinfo {author}
  {\bibfnamefont{M.~B.}\ \bibnamefont{Wise}},\ }%
  \bibfield{journal}{%
  \Doi{10.1016/0550-3213(82)90138-9}{\bibinfo {journal} {Nucl. Phys.}}\ }%
  \textbf{\bibinfo {volume} {B207}},\ \bibinfo {pages} {96} (\bibinfo {year}
  {1982})%
  \bibAnnoteFile{NoStop}{AlvarezGaume:1981wy}%
\bibitem{Dine:1981gu}%
  \BibitemOpen
  \bibfield{author}{%
  \bibinfo {author} {\bibfnamefont{M.}~\bibnamefont{Dine}}\ and\ \bibinfo
  {author} {\bibfnamefont{W.}~\bibnamefont{Fischler}},\ }%
  \bibfield{journal}{%
  \Doi{10.1016/0370-2693(82)91241-2}{\bibinfo {journal} {Phys. Lett.}}\ }%
  \textbf{\bibinfo {volume} {B110}},\ \bibinfo {pages} {227} (\bibinfo {year}
  {1982})%
  \bibAnnoteFile{NoStop}{Dine:1981gu}%
\bibitem{Dine:1982zb}%
  \BibitemOpen
  \bibfield{author}{%
  \bibinfo {author} {\bibfnamefont{M.}~\bibnamefont{Dine}}\ and\ \bibinfo
  {author} {\bibfnamefont{W.}~\bibnamefont{Fischler}},\ }%
  \bibfield{journal}{%
  \Doi{10.1016/0550-3213(82)90194-8}{\bibinfo {journal} {Nucl. Phys.}}\ }%
  \textbf{\bibinfo {volume} {B204}},\ \bibinfo {pages} {346} (\bibinfo {year}
  {1982})%
  \bibAnnoteFile{NoStop}{Dine:1982zb}%
\bibitem{Dine:1993yw}%
  \BibitemOpen
  \bibfield{author}{%
  \bibinfo {author} {\bibfnamefont{M.}~\bibnamefont{Dine}}\ and\ \bibinfo
  {author} {\bibfnamefont{A.~E.}\ \bibnamefont{Nelson}},\ }%
  \bibfield{journal}{%
  \Doi{10.1103/PhysRevD.48.1277}{\bibinfo {journal} {Phys. Rev.}}\ }%
  \textbf{\bibinfo {volume} {D48}},\ \bibinfo {pages} {1277} (\bibinfo {year}
  {1993}),\ \Eprint{http://arxiv.org/abs/hep-ph/9303230}{arXiv:hep-ph/9303230}%
  \bibAnnoteFile{NoStop}{Dine:1993yw}%
\bibitem{Dine:1994vc}%
  \BibitemOpen
  \bibfield{author}{%
  \bibinfo {author} {\bibfnamefont{M.}~\bibnamefont{Dine}}, \bibinfo {author}
  {\bibfnamefont{A.~E.}\ \bibnamefont{Nelson}},\ and\ \bibinfo {author}
  {\bibfnamefont{Y.}~\bibnamefont{Shirman}},\ }%
  \bibfield{journal}{%
  \Doi{10.1103/PhysRevD.51.1362}{\bibinfo {journal} {Phys. Rev.}}\ }%
  \textbf{\bibinfo {volume} {D51}},\ \bibinfo {pages} {1362} (\bibinfo {year}
  {1995}),\ \Eprint{http://arxiv.org/abs/hep-ph/9408384}{arXiv:hep-ph/9408384}%
  \bibAnnoteFile{NoStop}{Dine:1994vc}%
\bibitem{Dine:1995ag}%
  \BibitemOpen
  \bibfield{author}{%
  \bibinfo {author} {\bibfnamefont{M.}~\bibnamefont{Dine}}, \bibinfo {author}
  {\bibfnamefont{A.~E.}\ \bibnamefont{Nelson}}, \bibinfo {author}
  {\bibfnamefont{Y.}~\bibnamefont{Nir}},\ and\ \bibinfo {author}
  {\bibfnamefont{Y.}~\bibnamefont{Shirman}},\ }%
  \bibfield{journal}{%
  \Doi{10.1103/PhysRevD.53.2658}{\bibinfo {journal} {Phys. Rev.}}\ }%
  \textbf{\bibinfo {volume} {D53}},\ \bibinfo {pages} {2658} (\bibinfo {year}
  {1996}),\ \Eprint{http://arxiv.org/abs/hep-ph/9507378}{arXiv:hep-ph/9507378}%
  \bibAnnoteFile{NoStop}{Dine:1995ag}%
\bibitem{Meade:2008wd}%
  \BibitemOpen
  \bibfield{author}{%
  \bibinfo {author} {\bibfnamefont{P.}~\bibnamefont{Meade}}, \bibinfo {author}
  {\bibfnamefont{N.}~\bibnamefont{Seiberg}},\ and\ \bibinfo {author}
  {\bibfnamefont{D.}~\bibnamefont{Shih}},\ }%
  \bibfield{journal}{%
  \Doi{10.1143/PTPS.177.143}{\bibinfo {journal} {Prog. Theor. Phys. Suppl.}}\
  }%
  \textbf{\bibinfo {volume} {177}},\ \bibinfo {pages} {143} (\bibinfo {year}
  {2009}),\ \Eprint{http://arxiv.org/abs/0801.3278}{arXiv:0801.3278 [hep-ph]}%
  \bibAnnoteFile{NoStop}{Meade:2008wd}%
\bibitem{Dolgov:2002wy}%
  \BibitemOpen
  \bibfield{author}{%
  \bibinfo {author} {\bibfnamefont{A.}~\bibnamefont{Dolgov}},\ }%
  \bibfield{journal}{%
  \Doi{10.1016/S0370-1573(02)00139-4}{\bibinfo {journal} {Phys.Rept.}}\ }%
  \textbf{\bibinfo {volume} {370}},\ \bibinfo {pages} {333} (\bibinfo {year}
  {2002}),\ \Eprint{http://arxiv.org/abs/hep-ph/0202122}{arXiv:hep-ph/0202122
  [hep-ph]}%
  \bibAnnoteFile{NoStop}{Dolgov:2002wy}%
\bibitem{Bowen:2001in}%
  \BibitemOpen
  \bibfield{author}{%
  \bibinfo {author} {\bibfnamefont{R.}~\bibnamefont{Bowen}}, \bibinfo {author}
  {\bibfnamefont{S.~H.}\ \bibnamefont{Hansen}}, \bibinfo {author}
  {\bibfnamefont{A.}~\bibnamefont{Melchiorri}}, \bibinfo {author}
  {\bibfnamefont{J.}~\bibnamefont{Silk}},\ and\ \bibinfo {author}
  {\bibfnamefont{R.}~\bibnamefont{Trotta}},\ }%
  \bibfield{journal}{%
  \Doi{10.1046/j.1365-8711.2002.05570.x}{\bibinfo {journal}
  {Mon.Not.Roy.Astron.Soc.}}\ }%
  \textbf{\bibinfo {volume} {334}},\ \bibinfo {pages} {760} (\bibinfo {year}
  {2002}),\
  \Eprint{http://arxiv.org/abs/astro-ph/0110636}{arXiv:astro-ph/0110636
  [astro-ph]}%
  \bibAnnoteFile{NoStop}{Bowen:2001in}%
\bibitem{Cuoco:2005qr}%
  \BibitemOpen
  \bibfield{author}{%
  \bibinfo {author} {\bibfnamefont{A.}~\bibnamefont{Cuoco}}, \bibinfo {author}
  {\bibfnamefont{J.}~\bibnamefont{Lesgourgues}}, \bibinfo {author}
  {\bibfnamefont{G.}~\bibnamefont{Mangano}},\ and\ \bibinfo {author}
  {\bibfnamefont{S.}~\bibnamefont{Pastor}},\ }%
  \bibfield{journal}{%
  \Doi{10.1103/PhysRevD.71.123501}{\bibinfo {journal} {Phys.Rev.}}\ }%
  \textbf{\bibinfo {volume} {D71}},\ \bibinfo {pages} {123501} (\bibinfo {year}
  {2005}),\
  \Eprint{http://arxiv.org/abs/astro-ph/0502465}{arXiv:astro-ph/0502465
  [astro-ph]}%
  \bibAnnoteFile{NoStop}{Cuoco:2005qr}%
\bibitem{Lewis:1999bs}%
  \BibitemOpen
  \bibfield{author}{%
  \bibinfo {author} {\bibfnamefont{A.}~\bibnamefont{Lewis}}, \bibinfo {author}
  {\bibfnamefont{A.}~\bibnamefont{Challinor}},\ and\ \bibinfo {author}
  {\bibfnamefont{A.}~\bibnamefont{Lasenby}},\ }%
  \bibfield{journal}{%
  \Doi{10.1086/309179}{\bibinfo {journal} {Astrophys. J.}}\ }%
  \textbf{\bibinfo {volume} {538}},\ \bibinfo {pages} {473} (\bibinfo {year}
  {2000}),\
  \Eprint{http://arxiv.org/abs/astro-ph/9911177}{arXiv:astro-ph/9911177}%
  \bibAnnoteFile{NoStop}{Lewis:1999bs}%
\bibitem{Dutta:2009ix}%
  \BibitemOpen
  \bibfield{author}{%
  \bibinfo {author} {\bibfnamefont{S.}~\bibnamefont{Dutta}}, \bibinfo {author}
  {\bibfnamefont{S.~D.~H.}\ \bibnamefont{Hsu}}, \bibinfo {author}
  {\bibfnamefont{D.}~\bibnamefont{Reeb}},\ and\ \bibinfo {author}
  {\bibfnamefont{R.~J.}\ \bibnamefont{Scherrer}},\ }%
  \bibfield{journal}{%
  \Doi{10.1103/PhysRevD.79.103504}{\bibinfo {journal} {Phys. Rev.}}\ }%
  \textbf{\bibinfo {volume} {D79}},\ \bibinfo {pages} {103504} (\bibinfo {year}
  {2009}),\ \Eprint{http://arxiv.org/abs/0902.4699}{arXiv:0902.4699
  [astro-ph.CO]}%
  \bibAnnoteFile{NoStop}{Dutta:2009ix}%
\bibitem{Dutta:2010kb}%
  \BibitemOpen
  \bibfield{author}{%
  \bibinfo {author} {\bibfnamefont{S.}~\bibnamefont{Dutta}}\ and\ \bibinfo
  {author} {\bibfnamefont{R.~J.}\ \bibnamefont{Scherrer}},\ }%
  \bibfield{journal}{%
  \Doi{10.1103/PhysRevD.82.043526}{\bibinfo {journal} {Phys. Rev.}}\ }%
  \textbf{\bibinfo {volume} {D82}},\ \bibinfo {pages} {043526} (\bibinfo {year}
  {2010}),\ \Eprint{http://arxiv.org/abs/1004.3295}{arXiv:1004.3295
  [astro-ph.CO]}%
  \bibAnnoteFile{NoStop}{Dutta:2010kb}%
\bibitem{Doroshkevich:1989bf}%
  \BibitemOpen
  \bibfield{author}{%
  \bibinfo {author} {\bibfnamefont{A.~G.}\ \bibnamefont{Doroshkevich}},
  \bibinfo {author} {\bibfnamefont{M.}~\bibnamefont{Khlopov}},\ and\ \bibinfo
  {author} {\bibfnamefont{A.~A.}\ \bibnamefont{Klypin}},\ }%
  \bibfield{journal}{%
  \bibinfo {journal} {Mon. Not. Roy. Astron. Soc.}\ }%
  \textbf{\bibinfo {volume} {239}},\ \bibinfo {pages} {923} (\bibinfo {year}
  {1989})%
  \bibAnnoteFile{NoStop}{Doroshkevich:1989bf}%
\end{thebibliography}%
\end{document}